\documentclass{ifacconf}

\usepackage{natbib}        
\usepackage{amsmath, amssymb}
\usepackage{color,soul}
\usepackage{graphicx}      
\usepackage{lipsum}
\usepackage{xcolor}
\usepackage{bm}

\usepackage{mathtools} 
\usepackage{physics}
\usepackage{booktabs} 
\usepackage{bbm}
\usepackage{etoolbox}   
\usepackage{verbatim}   

\newcommand\scalemath[2]{\scalebox{#1}{\mbox{\ensuremath{\displaystyle #2}}}}
\newcommand\autoscalealign[1]{\resizebox{1\linewidth}{!}{\begin{minipage}{\linewidth} #1 \end{minipage}}}

\newtheorem{theorem}{Theorem}
\newtheorem{lemma}{Lemma}
\newtheorem{corollary}{Corollary}
\newtheorem{proposition}{Proposition}

\newtheorem{definition}{Definition}

\makeatletter
\def\endfigure{\end@float}
\def\endtable{\end@float}
\makeatother
\let\ifacconfcaptionwidth\captionwidth
\usepackage[caption=false]{subfig}
\let\captionwidth\ifacconfcaptionwidth

\newcommand\blfootnote[1]{%
  \begingroup
  \renewcommand\thefootnote{}\footnote{#1}%
  \addtocounter{footnote}{-2}%
  \endgroup
}

\newbool{extended}
\newbool{arxiv}
\setbool{extended}{true}
\setbool{arxiv}{true}

\ifboolexpr{bool {arxiv} and not bool {extended}}{\setbool{arxiv}{false}}{}

\newenvironment{extendedonly}{}{}
\ifboolexpr{not bool {extended}}{\AtBeginEnvironment{extendedonly}{\comment}%
\AtEndEnvironment{extendedonly}{\endcomment}}{}
\newenvironment{shortonly}{}{}
\ifbool{extended}{\AtBeginEnvironment{shortonly}{\comment}%
\AtEndEnvironment{shortonly}{\endcomment}}{}

\begin{document}
\begin{frontmatter}

\title{Nonlinear MPC for Offset-Free Tracking of systems learned by GRU Neural Networks}

\author{Fabio Bonassi\thanksref{*}} 
\author{Caio Fabio Oliveira da Silva\thanksref{*}} 
\author{Riccardo Scattolini}

\thanks[*]{Indicates equal contribution to this work.}

\address{Politecnico di Milano, Dipartimento di Elettronica, Informazione e Bioingegneria, Via Ponzio 34/5, 20133 Milano, Italy \\
	(e-mail: name.surname@polimi.it)}

\begin{abstract}
	The use of Recurrent Neural Networks (RNNs) for system identification has recently gathered increasing attention, thanks to their black-box modeling capabilities.
	Albeit RNNs have been fruitfully adopted in many applications, only few works are devoted to provide rigorous theoretical foundations that justify their use for control purposes.
	The aim of this paper is to describe how stable Gated Recurrent Units (GRUs), a particular RNN architecture, can be trained and employed in a Nonlinear MPC framework to perform offset-free tracking of constant references with guaranteed closed-loop stability. 
	The proposed approach is tested on a \emph{pH} neutralization process benchmark, showing remarkable performances.
\end{abstract}

\begin{keyword}
Machine Learning, Nonlinear Model Predictive Control, Model and Identification of Nonlinear Systems, Offset-free Tracking
\end{keyword}

\end{frontmatter}

\section{Introduction} \label{sec:introduction}  

In recent years, the advances in machine learning techniques have motivated the control systems community to investigate the use of Recurrent Neural Networks (RNNs) for the identification of nonlinear dynamical systems.
Among the various RNN architectures,  Gated Recurrent Units (GRUs, \cite{chung2014empirical}) and Long Short-Term Memory (LSTM, \cite{hochreiter1997long}) networks have proven to be particularly suited for system identification tasks \citep{forgione2020model}, as their stateful structure is able to retain long-term memory of past inputs and states.
\ifbool{arxiv}{\blfootnote{\hspace{-4mm} © 2021 the authors. This work is the extended version of the paper accepted at the Third IFAC Conference on Modelling, Identification and Control of Nonlinear Systems (MICNON 2021) for publication under a Creative Commons Licence CC-BY-NC-ND.}}{}

In light of their superior modeling capabilities, these gated RNNs have been widely adopted by control practitioners, in conjunction with nonlinear model-based control strategies, such as Model Predictive Control (MPC), especially in process control domain \citep{wong2018recurrent}.
For example, in \cite{terzi2020learning} a complex nonlinear system is identified by means of an LSTM network, which is then used as a prediction model in a Nonlinear MPC (NMPC) regulator.
A similar strategy is adopted in \cite{lanzetti2019recurrent}, where a GRU network is used to identify and control a paper machine.

Despite the adoption of RNNs in many data-driven control applications, only limited theoretical foundations are nowadays available to justify the use of these NN architectures in systems control domain.
The available theoretical results are mainly focused on achieving provably stable models.
In \cite{bonassi2020stability} and \cite{stipanovic2020stability}, the authors derive conditions that should be satisfied by network's weights in order to guarantee its stability.
Similar conditions have been derived for LSTMs networks in \cite{bonassi2019lstm} and \cite{terzi2021learning}, where an NMPC control strategy with guaranteed closed-loop stability is also formulated.

While the stability of RNN architectures can be used to design standard closed-loop-stable MPC laws, see \cite{terzi2021learning}, the tracking performances of these control strategies mainly depend on the accuracy of the trained networks.
To address this problem, in the literature several offset-free NMPC formulations have been proposed, which allow to achieve perfect tracking of output references with stability guarantees \citep{pannocchia2015offset}.
In \cite{morari2012} offset-free tracking is achieved by enlarging the system with a disturbance model.
A state observer is then designed for the enlarged system, so that the disturbance is estimated and compensated by the NMPC control system.
Another approach is the one proposed by \cite{magni2001output}, where the system is augmented with an integral action on the output tracking error, a state observer is designed and a stabilizing NMPC law is synthesized for the augmented system.

The aim of this work is to further elaborate on \cite{bonassi2020stability} and \cite{terzi2021learning}, and to describe an unitary approach for the identification and off-set free control of an unknown dynamical system.
Specifically, we herein propose to use a stable GRU network to identify the unknown plant from data collected from the real system.
Once trained, this network is used as predictive model for an NMPC controller which ensures offset-free tracking of constant references with closed-loop stability guarantees, adopting the control strategy proposed by \cite{magni2001output}.
Therefore, sufficient conditions under which it is possible to design a converging state observer for the augmented system are devised.
The stabilizing NMPC law is thus formulated for the enlarged system.
The approach is tested on a \emph{pH} neutralization process benchmark system, showing remarkable results in terms of reference tracking capabilities under control constraints.

\subsection{Notation} \label{sec:notation } 
If $v$ is a vector, $v^\prime$ is its transpose and $\| v \|$ its Euclidean norm.
For simplicity, vectors whose components are all equal to one and to zero are compactly denoted as $1$ and $0$.
By boldface fonts we indicate sequences of vectors, i.e.  $\bm{v} = \{ v(0), v(1), ... \}$, and $\| \bm{v} \|_p = \max_{k \geq 0} \| v(k) \|_p$.
For compactness, for time-varying quantities the generic time index $k$ is omitted if it can be inferred from the context.
Superscript $^+$ indicates some quantity at the successive time instant, i.e. at $k+1$. Therefore, $x = x(k)$ and $x^+ = x(k+1)$.
The Hadamard product between matrices or vectors is denoted by $\circ$.
By $\sigma$ and $\phi$ we indicate the sigmoid and tanh activation functions, respectively, i.e. $\sigma(x) = \frac{1}{1+e^{-x}}$ and $\phi(x) = \text{tanh}(x)$.

\section{Model} \label{sec:model}
In this section we introduce GRUs, discussing how these networks can be used to identify nonlinear dynamical systems and how stability properties can be enforced. 
For simplicity we focus on single-layer GRU networks, but the approach can be easily extended to multi-layer networks.

Let us now introduce a single-layer GRU expressed in state-space form \citep{bonassi2020stability}
 \begin{equation} \label{eq:model:gru_full}
 \scalemath{0.95}{
 \Sigma: \,\,
 	\left\{
 	\begin{array}{l}
 		x^+ = z \circ x + (1-z) \circ \phi(W_r u + U_r \, f \circ x + b_r) \\
 		z(x, u) = \sigma(W_z u + U_z x + b_z) \\
 		f(x, u) = \sigma(W_f u + U_f x + b_f) \\
 		y = U_o x + b_o
 	\end{array}
 	\right.,}
 \end{equation}
 where $x \in \mathbb{R}^n$ is the state, $u \in \mathbb{R}^m$ is the input, $y \in \mathbb{R}^p$ is the output, and $m=p$ is assumed.
 Functions $z = z(x, u)$ and $f = f(x, u)$ are known as update and forget gates, respectively.
 The matrices $W_\star$, $U_\star$, $b_\star$, are the weights of the network, and must be tuned during the training procedure in such a way that \eqref{eq:model:gru_full} approximates sufficiently well the unknown plant from which the data was collected. 
 For compactness, $\Sigma$ may be re-written as
 \begin{equation} \label{eq:model:gru}
     \Sigma: \, \left\{ \begin{array}{l}
     x^+ = \varphi(x, u) \\
     y = \eta(x)
     \end{array} \right.,
 \end{equation}
 where $\varphi(x, u)$ and $\eta(x)$ can be easily retrieved from \eqref{eq:model:gru_full}.
 
 In the following we assume that the input $u$ is unity-bounded, i.e. $u \in \mathcal{U} = [-1, 1]^m$ or, equivalently, $\| u \|_\infty \leq 1$.
 Note that this is a quite customary assumption when working with neural networks, and it can be easily satisfied by means of a suitable normalization \citep{terzi2021learning}. 
 It can be then shown that in finite time the state of the GRU enters a unity-boxed invariant set.
 
 \begin{lemma}[\cite{bonassi2020stability}] \label{lemma:invset}
 	For any initial state $\bar{x}$ and any input sequence $\bm{u}$, there exists a finite time instant $\bar{k} \geq 0$ at which the state enters its invariant set $\mathcal{X} = [-1, 1]^n$, i.e. $x(k) \in \mathcal{X} \, \forall k \geq \bar{k}$.
 \end{lemma}
 
 We remind that some function $\psi(s)$ is said to be a $\mathcal{K}_\infty$ function if $\psi(0) = 0$, it is strictly increasing and $\psi(s) \to \infty$ as $s \to +\infty$. A function  $\psi(s, t)$ is of class $\mathcal{KL}$ if it is $\mathcal{K}_\infty$ with respect to $s$ and, in addition, $\psi(s, t) \to 0$ as  $t \to  0$.
  Under these premises, we can summarize the main stability results for GRU networks. 
  For more details, the interested reader is addressed to \cite{bonassi2020stability}.
  
  \begin{definition}[$\delta$ISS]
	System \eqref{eq:model:gru} is said to be Incrementally Input-to-State Stable ($\delta$ISS) if there exists functions $\beta \in \mathcal{KL}$ and $\gamma \in \mathcal{K}_\infty$ such that, for any pair of initial states $\bar{x}_a$ and $\bar{x}_b$, and any pair of input sequences $\bm{u}_a$ and $\bm{u}_b$ it holds that
	\begin{equation}
	\begin{aligned}
		&\| x(k, \bar{x}_a, \bm{u}_a) -  x(k, \bar{x}_b, \bm{u}_b) \| \qquad \qquad \\
		& \qquad \leq \beta(\|\bar{x}_a - \bar{x}_b \|, k) + \gamma(\| \bm{u}_a - \bm{u}_b \|),
	\end{aligned}
	\end{equation}
	where $x(k, \bar{x}, \bm{u})$ is the state trajectory at time $k$, when \eqref{eq:model:gru} is initialized in $\bar{x}$ and it is feed by the input sequence $\bm{u}$.
  \end{definition}
  
  Note that, since $\beta \in \mathcal{KL}$, this stability property implies that the effects of initial conditions asymptotically vanish, guaranteeing that the modeling performances of the network is independent of the initialization.
   
  As $\delta$ISS plays a key role in the proposed control design, see Section \ref{sec:control}, the following sufficient condition for the $\delta$ISS of GRU networks proposed by \cite{bonassi2020stability} is enforced during the training of the network.
\begin{theorem}[\cite{bonassi2020stability}] \label{thm:deltaiss}
	A sufficient condition for the $\delta$ISS of system \eqref{eq:model:gru} is that
	\begin{equation} \label{eq:model:deltaiss_condition}
	\scalemath{0.95}{
		\| U_r \|_\infty \Big(\frac{1}{4} \| U_f \|_\infty + \bar{\sigma}_f \Big) < 1 - \frac{1}{4} \frac{1 + \bar{\phi}_r}{1 - \bar{\sigma}_z} \| U_z \|_\infty,}
	\end{equation} 	
	where $\bar{\sigma}_z$, $\bar{\phi}_r$, and $\bar{\sigma}_f$ are defined as
	\begin{equation} \label{eq:model:sigma_bar}
	\scalemath{0.925}{
		\begin{aligned}
			\bar{\sigma}_z &= \sigma (\| W_z \quad U_z \quad b_z \|_\infty), \\
			\bar{\phi}_r &= \phi (\| W_r \quad U_r \quad b_r \|_\infty), \\
			\bar{\sigma}_f &= \sigma (\| W_f \quad U_f \quad b_f \|_\infty).
		\end{aligned}}
	\end{equation}
\end{theorem}
 
Note that this sufficient condition boils down to an inequality on the weights of the network.
The fulfillment of \eqref{eq:model:deltaiss_condition} can be enforced as a constraint during the training procedure of the network.
In case of use of unconstrained training algorithms, this condition may be relaxed, penalizing the violation of \eqref{eq:model:deltaiss_condition} in the loss function, as outlined in Section \ref{sec:numerical} and discussed in \cite{bonassi2020stability}.

\section{Control architecture design} \label{sec:control}

Thus far, a black-box model with guaranteed stability properties has been presented.
In this Section, we assume that the $\Sigma$ has been trained to model the unknown plant, and that the weights of this network satisfy condition \eqref{eq:model:deltaiss_condition}, so that  Theorem \ref{thm:deltaiss} guarantees the $\delta$ISS of the network.

The goal of this section is to employ such model for offset-free tracking of constant reference signals.
In particular, we adopt the control strategy proposed by \cite{magni2001output} which, informally speaking, consists in the following steps: \emph{(i)} augmenting the system model with an integrator on the output tracking error, \emph{(ii)} designing a weak detector for the augmented plant, and \emph{(iii)} synthesizing a stabilizing NMPC law for the augmented system.

\begin{figure}[t]
    \centering
    \includegraphics[width=0.9 \linewidth]{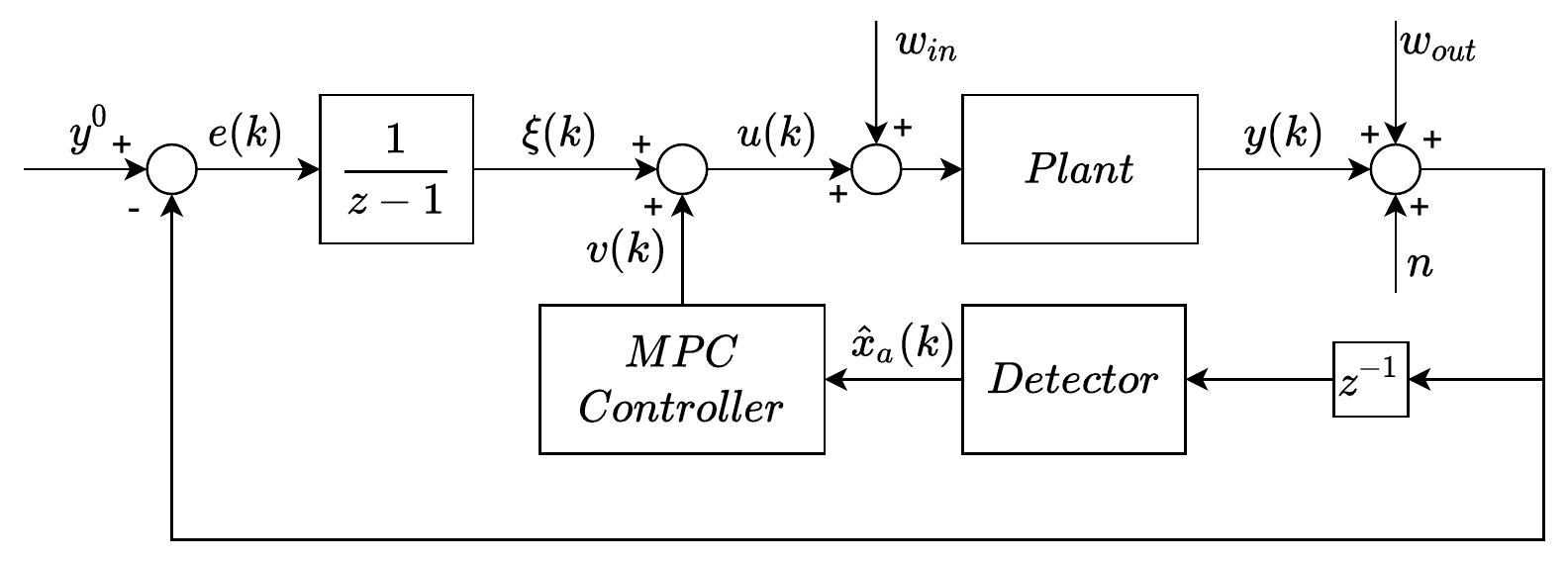}
    \caption{A representation of the controller architecture.}
    \label{fig:controller_scheme}
\end{figure}

The resulting control architecture is depicted in Figure \ref{fig:controller_scheme}.
Under mild assumptions on the model's smoothness, and provided that the state observer is a weak detector for the augmented system and that the NMPC law stabilizes the reference equilibrium, \cite{magni2001output} guarantees offset-free tracking of piece-wise constant references.

\begin{rem}
Differently from \cite{magni2001output}, in general $(x^0, u^0, y^0) = (0, 0, 0)$ is not an equilibrium of $\Sigma$, i.e. $\varphi(0, 0) \neq 0$ and $\eta(0) \neq 0$.
However, since the designed observer is a weak detector with respect to any equilibrium, it is enough to design the NMPC law stabilizing the generic equilibrium $(x^0, u^0, y^0)$ of $\Sigma$.
\end{rem}

\subsection{Model augmentation and state observer design}
To achieve offset-free tracking of the output reference $y^0 \in \mathbb{R}^p$, the system model is augmented with the discrete-time integral of the tracking error.
The enlarged system $\Sigma_a$ reads as
\begin{equation} \label{eq:control:augmented1}  
    \Sigma_a: \,\,
    \left\{
    \begin{array}{l}
         x^+ = \varphi(x, u)  \\
         y = \eta(x) \\
         \xi^+ = \xi + y^0 - y \\
         u = v + \xi
    \end{array}
    \right.,
\end{equation}
where $\xi \in \mathbb{R}^p$ is the state of the integrator, and $v$ is the exogenous input computed by the NMPC.
Denoting by $x_a = [ x^{\prime}, \xi^{\prime}]^{\prime}$ the state of the augmented system and by $y_a = [y^\prime, \xi^\prime]^\prime$ its output, \eqref{eq:control:augmented1} can be compactly denoted by 
\begin{equation} \label{eq:control:augmented_simple}
    \Sigma_a: \,\, \left\{
    \begin{array}{l}
         x_a^+ = \varphi_a(x_a, v, y^0)  \\
         y_a = \eta_a(x_a, y^0)
    \end{array}
    \right..
\end{equation}

Then, a state observer needs to be designed for the enlarged system.
This observer is required to be a weak detector of \eqref{eq:control:augmented_simple} for the equilibrium $x^0_a = [ x^{0 \prime}, u^{0 \prime}]^\prime$, $v^0 = 0$, and $y^0_a = [y^{0 \prime}, u^{0 \prime}]^\prime$, in the sense specified by the following definition \citep{magni2001bookchapter}.

\begin{definition} \label{def:detector}
    System \eqref{eq:control:augmented_simple} is weakly detectable with respect to the equilibrium $(x^0_a, 0, y^0_a)$ if there exists a function $g: \mathbb{R}^{n+p} \times \mathbb{R}^m \times \mathbb{R}^p \times \mathbb{R}^{2p} \rightarrow \mathbb{R}^{n+p}$ and a function $V_e \in \mathcal{C}^1$ such that the following properties hold:
\begin{enumerate}
    \item $g(\hat{x}_a, v, y^0, y_a)$ is of class $\mathcal{C}^1$ and $\hat{x}_a^0 = g(\hat{x}_a^0, 0, y^0, y_a^0)$
    \item There exist functions $a_e$, $b_e$, and $c_e$ of class $\mathcal{K}_\infty$ such that, for any $x, \hat{x} \in \mathcal{X}$, any $\xi$ and $v$ such that $\| \xi + v \|_\infty \leq 1$, and any $\hat{\xi}$ such that $\| \hat{\xi} + v \|_\infty \leq 1$, it holds 
    \begin{subequations}
    \begin{gather}
        a_e(\| x_a - \hat{x}_a \|) \leq V_e(x_a, \hat{x}_a) \leq b_e(\| x_a - \hat{x}_a \|), \\
        V_e(x_a^+, \hat{x}_a^+) - V_e(x_a, \hat{x}_a)\leq -c_e(\| x_a - \hat{x}_a  \|),
    \end{gather}
    \end{subequations}
    where $x_a^+ = \varphi_a(x_a, v, y^0)$ and $\hat{x}_a^+ = g(\hat{x}_a, v, y^0, y_a)$.
\end{enumerate}
Then, $\hat{x}^+_a = g(\hat{x}_a, v, y^0, y_a)$ is called \emph{weak detector} of $\Sigma_a$.
\end{definition}

The proposed state observer for the augmented system reads as follows
 \begin{equation} \label{eq:control:detector}
 	\scalemath{0.8}{
 	\mathcal{O}_a:  \left\{
 	\begin{array}{l}
 		\hat{x}^+ = \hat{z} \circ \hat{x} + (1-\hat{z}) \circ \phi(W_r (v + \hat{\xi}) + U_r \, \hat{f} \circ \hat{x} + b_r) \\
 		\hat{z}(\hat{x}_a, v) = \sigma(W_z (v + \hat{\xi})  + U_z \hat{x} + b_z + L_{z \xi} (\xi - \hat{\xi}) + L_{zy} (y-\hat{y})) \\
 		\hat{f}(\hat{x}_a, v) = \sigma(W_f (v + \hat{\xi})  + U_f \hat{x} + b_f + L_{f \xi} (\xi - \hat{\xi}) + L_{fy} (y - \hat{y})) \\
 		\hat{\xi}^+ = \hat{\xi} + y^0 - \hat{y} + L_{\xi y} ( y - \hat{y}) + L_{\xi \xi} (\xi - \hat{\xi}) \\
 		\hat{y} = U_o \hat{x} + b_o 
 	\end{array}
 	\right.}
 \end{equation}

The following results allow to provide guidelines for the tuning of the observer gains $L_\star$ so as to guarantee that $\mathcal{O}_a$ is a weak detector of $\Sigma_a$.
\ifbool{extended}{The proofs are reported in the Appendix.}{The proofs are reported in the extended version of this paper \citep{bonassi2021extended}.}

\begin{theorem} \label{thm:observer}
    A sufficient condition under which $\mathcal{O}_a$ is a weak detector of system \eqref{eq:control:augmented_simple}, in the sense specified by Definition \ref{def:detector}, is that there exists $\delta > 0$ such that
    \begin{subequations}
    \begin{equation} \label{eq:control:observer_theorem:xx}
    \scalemath{0.825}{
        \| U_r \|_\infty \Big(\frac{1}{4} \| U_f - L_{fy} U_o \|_\infty + \bar{\sigma}_f \Big) + \frac{1}{4} \frac{1 + \bar{\phi}_r}{1 - \bar{\sigma}_z} \| U_z - L_{zy} U_o \|_\infty \leq 1 - \delta}
    \end{equation}
    and that the matrix $A_\delta$ defined as
    \begin{equation} \label{eq:control:observer_theorem:A}
        A_\delta= \begin{bmatrix}
        1 - \delta &&  \alpha\\
        \| U_o \|_\infty \| I + L_{\xi y} \|_\infty & \,\,\,& \| I - L_{\xi \xi} \|_\infty
        \end{bmatrix}
    \end{equation}
    is Schur stable, where 
    \begin{equation} \label{eq:control:observer_theorem:a12}
    \scalemath{0.825}{
        \alpha = \frac{1}{4} \| W_z - L_{z\xi} \|_\infty ( 1 + \| W_r \|_\infty + \frac{1}{4} \| U_r \|_\infty \| W_f -L_{f \xi} \|_\infty ).}
    \end{equation}
    \end{subequations}
\end{theorem}

\begin{proposition} \label{prop:obsv_tuning}
    The observer $\mathcal{O}_a$ satisfies Theorem \ref{thm:observer} if its gains $L_\star$ are a feasible solution of the following optimization problem
    \begin{equation} \label{eq:control:observer_synthesis}
    \scalemath{0.9}{
        \begin{aligned}
            \min_{L_{\star}} \,\, & \| A_\delta || \\
            \text{s.t.} \,\,\, &  \eqref{eq:control:observer_theorem:xx} \\
            & \delta ( 1 - \| I - L_{\xi \xi} \|_\infty ) > \alpha \| U_0 \|_\infty \| I + L_{\xi y} \|_\infty \\
            & (1-\delta) \| I - L_{\xi \xi} \|_\infty < 1 +  \alpha \| U_0 \|_\infty \| I + L_{\xi y} \|_\infty \\
            & \delta > 0 
        \end{aligned}}
    \end{equation}
\end{proposition}

Note that Proposition \ref{prop:obsv_tuning} involves solving a non-linear optimization problem to spot suitable weights of $\mathcal{O}_a$.
The constraints of \eqref{eq:control:observer_synthesis} enforce the nominal convergence of the observed states $\hat{x}_a$ to the real states $x_a$. 
Furthermore, recalling that the spectral radius of the matrix is bounded by its norm, i.e. $\rho(A_\delta) < \| A_\delta \|$, by minimizing $\| A_\delta \|$ one minimizes the maximum eigenvalue, likely making the observer convergence faster.

\begin{corollary} \label{cor:trivial_observer}
    If the model satisfies the $\delta$ISS condition \eqref{eq:model:deltaiss_condition}, the optimization problem \eqref{eq:control:observer_synthesis} admits a feasible solution, corresponding to $L_{fy} = L_{zy} = 0_{n, p}$, 
    $L_{z\xi} = W_z$, $ L_{\xi \xi} < \lambda I_{p, p}$ with $\lambda  \in (0,1)$, and any $L_{f \xi}$ and $L_{\xi y}$.
\end{corollary}
Therefore, the $\delta$ISS property of the model allows to guarantee the weak detectability of the augmented system, and the designed observer $\mathcal{O}_a$ enjoys an estimation error nominally converging to zero.

\subsection{MPC control design}
Having designed a weak detector for the augmented system $\Sigma_a$, an NMPC stabilizing control law is now synthesized.
As discussed, we consider the equilibrium $(x_a^0, 0, y_a^0)$, such that $x_a^0 = \varphi_a(x_a^0, 0, y^0)$ and $y_a^0 = \eta_a(x_a^0, y^0)$.

Let $A_a$, $B_a$, and $C_a$ be the matrices of the linearized augmented system around this equilibrium
\begin{equation}
\scalemath{0.85}{
    A_a = \eval{\pdv{\varphi_a}{x_a}}_{(x_a^0,0, y^0)}\!\!, \,\,\, B_a = \eval{\pdv{\varphi_a}{v}}_{(x_a^0,0, y^0)}\!\!,  \,\,\, C_a = \eval{\pdv{\eta_a}{x_a}}_{(x_a^0, y^0)}}\!\!.
\end{equation}
Henceforth it is assumed that the triplet $(A_a, B_a, C_a)$ is stabilizable, observable and does not have transmission at point one in the complex plane. 
These conditions guarantee the existence of an open neighborhood of $y^0$ where any reference $\tilde{y}^0$ can define an equilibrium of the system (\cite{de1997narx}, Theorem 1).

According to the MPC approach, at each time-step a Finite Horizon Optimization Control Problem (FHOCP) is solved to spot the control sequence $\bm{v} = [v^\prime(0), ..., v^\prime(N_c -1)]^\prime$ minimizing a cost function $J$ throughout the horizon $\{1, ..., N_p\}$.
The parameter $N_p$ is called prediction horizon, and it is the time window throughout which the system is simulated and the cost function is evaluated.
In addition, we define the parameter $N_c \leq N_p$, called control horizon.
From the end of the control horizon onward, the control $v$ is assumed to be computed by the auxiliary law $v_{lq} = - K_{lq} (x_a - x_a^0)$, where $K_{lq}$ is the LQ$_\infty$ gain stabilizing the linearized system, i.e. $(A_a, B_a, C_a)$, and computed with the state and control weight matrices $Q \succ 0$ and $R \succ 0$.

According to the Receding Horizon technique, once the optimal solution of the FHOCP $\bm{v}^*$ is retrieved, the first optimal control move, $v^*(0)$, is applied to the system, i.e. $v(k) = v^*(0)$.
At the following time instant, the model is re-initialized in the observed state $\hat{x}_a$, and the entire procedure is repeated.

The FHOCP is stated as follows \\
\autoscalealign{
    \scriptsize
    \begin{subequations} \label{eq:control:nmpc}
    \begin{align}
        \min_{\bm{v}} \,\, & J(\bm{x}_a, \bm{v}) & \label{eq:control:nmpc:J}\\
        \text{s.t.} \,\, & x_a(0) = \hat{x}_a(k) \label{eq:control:nmpc:x_init} \\
        & \tilde{\xi}(0) = \xi(k) \label{eq:control:nmpc:int_init} \\
        & x_a(i+1) = \varphi_a(x_a(i), v(i), y^0)   & \forall i \in \{ 1, ...., N_p -1 \} \label{eq:control:nmpc:x_mpc} \\
        & v(i) = -K_{lq} (x_a(i) - x_a^0) &\forall i \in \{ N_c, ..., N_p - 1\} \label{eq:control:nmpc:aux} \\
        & \tilde{\xi}(i+1) = \tilde{\xi}(i) + y^0 - E_p y_a(i) & \forall i \in \{0, ..., N_p - 1\} \label{eq:control:nmpc:integrator}\\ 
        & \tilde{\xi}(i) + v(i) \in [-1, 1]  & \forall i \in \{0, ..., N_p - 1\}  \label{eq:control:nmpc:constraints}\\
        & x_a(N_p ) \in \Omega_\omega &  \label{eq:control:nmpc:terminal}
    \end{align}
    \end{subequations}
    \normalsize
}
where a standard quadratic cost function is adopted
\begin{equation}
\scalemath{0.85}{
\begin{aligned}
    J(\bm{x}_a, \bm{v}) =& \sum_{i=0}^{N_c-1} \big( \| x_a(i) - x_a^0 \|^2_Q + \| v(i) \|^2_R \big) \\
    & +\sum_{i=N_c}^{N_p-1} \| x_a(i) - x_a^0 \|^2_{{Q}_{lq}} + V_f(x_a(N_p)),
\end{aligned}}
\end{equation}
and $Q$ and $R$ are the matrices used to compute $K_{lq}$.
Note that after the control horizon $N_c$ the LQ$_\infty$ auxiliary control law \eqref{eq:control:nmpc:aux} is assumed to be adopted, thus $Q_{lq} = Q + K_{lq}^\prime R K_{lq}$.
The term $V_f(x_a(N_p))$ is the terminal cost, defined as a non-quadratic approximation of the LQ$_\infty$ cost-to-go, as discussed later this section.

The system model is embedded in the FHOCP by means of \eqref{eq:control:nmpc:x_mpc}, where the state $x_a$ is initialized in the values estimated by the weak detector $\mathcal{O}_a$ \eqref{eq:control:nmpc:x_init}.
Moreover, in addition to the integrator model embedded in $\Sigma_a$, a further integrator model with state $\tilde{\xi}$ is included. 
This integrator is initialized in the true (known) integrator state $\xi(k)$, see \eqref{eq:control:nmpc:int_init}, and it is used to ensure that the input sequence $v$ satisfies the input constraints \eqref{eq:control:nmpc:constraints} throughout the prediction horizon.
The state $\hat{\xi}$ is then updated according to the integrator model \eqref{eq:control:nmpc:integrator}, where $E_p$ is the selection matrix extracting $y$ from $y_a$, i.e. $E_p y_a =  y$.

Eventually, the terminal constraint \eqref{eq:control:nmpc:terminal} is imposed, by means of which the state at the end of the prediction horizon is forced to lie within the terminal set $\Omega_\omega$. 
A suitable design of the terminal cost $V_f(x_a(N_p))$ and of the terminal set $\Omega_\omega$ is paramount to guarantee the closed-loop stability and the recursive feasibility of the NMPC scheme. 
Herein we consider the approach proposed in \cite{magni2001stabilizing} for the computation of these terminal ingredients.

By definition, the terminal set $\Omega_\omega$ is the set of  states that can be steered to the equilibrium $(x_a^0, 0)$ by the stabilizing LQ$_\infty$ control law $v_{lq} = -K_{lq} (x_a - x_a^0)$. 
In practice, this set is approximated by an inner hyper-ellipsoid in which a Lyapunov function defined by $V_{l} = (x_a - x_a^0)^\prime \Pi (x_a - x_a^0)$ is decreasing, that is
\begin{equation} \label{eq:control:terminal}
\scalemath{0.85}{
    \begin{aligned}
        \Omega_\omega = \big\{ x_a& \in \mathbb{R}^{n + p} \,\, | \,\,  \| x_a - x_a^0 \|^2_\Pi \leq \omega, \\
        & v_{lq} = - K_{lq} (x_a - x_a^0) \in [-1, 1], \\
        & \| \varphi_a(x_a, v_{lq}, y^0) - x_a^0\|_\Pi^2 - \| x_a - x_a^0 \|_\Pi^2 + \| x_a - x_a^0 \|_{\gamma I}^2 \leq 0 \big\},
    \end{aligned}}
\end{equation}
where $\Pi$ is the solution of the LQ$_\infty$ Lyapunov equation $(A_{a}-B_{a} K_{lq})^\prime \Pi (A_{a}-B_{a} K_{lq}) - \Pi =  - \tilde{Q}$, with $\tilde{Q} \succ 0$ and $\gamma > 0$ being design parameters.

Concerning the terminal cost $V_f(x_a(N_p))$, we define it as a finite-step approximation of the cost required to steer the system from $\chi_a(0) = x_a(N_p)$ to the equilibrium $x_a^0$ by means of the LQ$_\infty$ auxiliary control law.
Specifically, being $\chi_a(j+1) = \varphi_a(\chi_a(j), -K_{lq}(\chi_a(j) - x_a^0), y^0)$ the state of the system subject to the auxiliary control law, the terminal cost can be approximated by
\begin{equation} \label{eq:control:terminal_cost}
    V_f(x_a(N)) = \sum_{j=0}^{N_{f}-1} || \chi(j) ||_{Q_{lq}}^2,
\end{equation}
where it is reminded that $Q_{lq} = Q + K_{lq}^\prime R K_{lq}$, and $N_f$ is the index that determines the truncation of the infinite horizon cost while ensuring an accurate approximation.

\begin{rem}
Let $v(k) = \kappa^{RH}(\hat{x}_a(k))$ be the Receding Horizon control law.
According to \cite{ohno1978new}, this control law is Lipschitz continuous, since the constraints of \eqref{eq:control:nmpc} are regular. 
Thus, provided that a weak detector such as \eqref{eq:control:detector} is employed, by Theorem 3 of \cite{magni2001output} the closed-loop system $x_a^+ = \varphi_a(x_a, \kappa^{RH}(\hat{x}_a), y^0)$ is guaranteed to converge to the target equilibrium $(x_a^0, 0, y_a^0)$, i.e. offset-free tracking is achieved.
\end{rem}

\section{Numerical results} \label{sec:numerical}

\begin{extendedonly}
    \begin{figure}[t]
        \centering
        \includegraphics[width=0.6\linewidth]{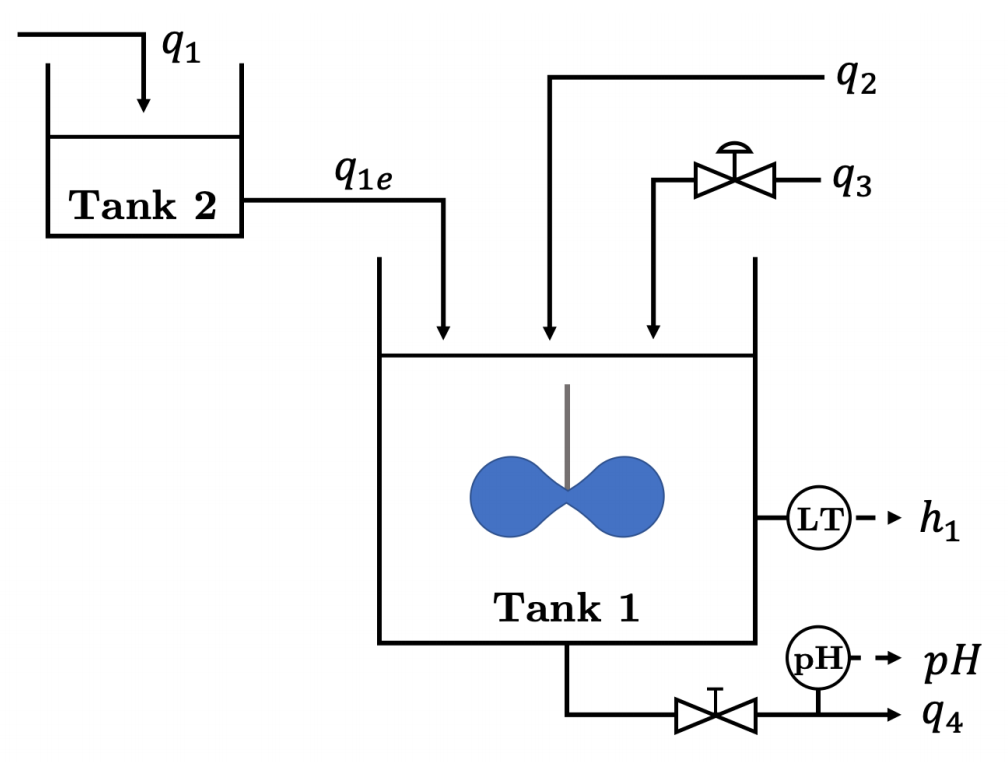}
        \caption{A representation of the pH neutralization process.}
        \label{fig:pH_scheme}
    \end{figure}
\end{extendedonly}

In this Section, the \emph{pH} neutralization process described by \cite{pHbenchmark} is used as a test-bed for evaluating both the modeling capabilities of GRU networks and the performance of the proposed control architecture.

As depicted in Figure \ref{fig:pH_scheme}, the \emph{pH} process consists of two tanks.
Tank 2 is fed by an acid flow rate $q_{1}$, and its output is $q_{1e}$.
Since hydraulic dynamics are faster than the others involved, it is assumed that $q_1 = q_{1e}$.
Tank 1, named reactor tank, receives three flows: the acid stream $q_{1e}$, a buffer flow $q_2$ an alkaline flow $q_3$. 
The only control variable is $q_3$, while the other two are considered as disturbances.
The goal is to set the \emph{pH} concentration of the output flow rate, $q_4$, to the desired level.
Hence, the process has only one input, $u=q_3$, and one output, $y = pH(q_4)$. 
The underlying SISO nonlinear model can be described by the following equations
\begin{equation} \label{eq:example:ph_process}
\scalemath{0.85}{
    \begin{aligned}
    \dot{x}(t) &= f_1(x(t)) + f_2(x(t))u(t) + f_3(x(t))d(t), \\
    0 &= c(x(t), y(t)).
    \end{aligned}}
\end{equation}
\begin{shortonly}
For space reasons, the exact expression of functions $f_1$, $f_2$, and $f_3$ is here omitted, and the interested reader is addressed to the extended version \citep{bonassi2021extended}.
\end{shortonly}
\begin{extendedonly}
where
\begin{equation} \label{eq:example:ph_full}
    \scalemath{0.75}{
    \begin{aligned}
        & \qquad \qquad \qquad f_{1}(x(t))=\begin{bmatrix}
        \frac {q_{1}}{A_{1}x_{3}}(W_{a1}-x_{1})\\
        \frac {q_{1}}{A_{1}x_{3}}(W_{b1}-x_{2})\\
        \frac {1}{A_{1}}(q_{1}-C_{v4}(x_{3}+z)^{n})
        \end{bmatrix}, \\
        &f_{2}(x(t))=\begin{bmatrix}
        \frac {1}{A_{1}x_{3}}(W_{a3}-x_{1})\\
        \frac {1}{A_{1}x_{3}}(W_{b3}-x_{2})\\
        \frac {1}{A_{1}}
        \end{bmatrix}, \quad
        f_{3}(x(t))=\begin{bmatrix}
        \frac {1}{A_{1}x_{3}}(W_{a2}-x_{1}) \\
        \frac {1}{A_{1}x_{3}}(W_{b2}-x_{2}) \\
        \frac {1}{A_{1}}
        \end{bmatrix}, \\
        &c(x,y)=x_{1}+10^{y-14}+10^{-y}+x_{2}\frac {1+2\cdot 10^{y-pK_{2}}}{1+10^{pK_{1}-y}+10^{y-pK_{2}}}.
    \end{aligned}}
\end{equation}
In Table \ref{tab:pH_parameters} parameters and nominal operating condition  of model \eqref{eq:example:ph_full} are reported.
 \begin{table}[t]
    \centering
    \caption{Nominal operating conditions of the \emph{pH} reactor plant.}
    \begin{tabular}{lll}
    \toprule
    z = 11.5 cm     & $W_{a1}$ = 3e3     & $W_{b4}$ = 5.28e2  \\
    $C_{v4}$ = 4.59 & $W_{b1}$ = 0       & $q_1$ = 16.6mL/s   \\
    n = 0.607       & $W_{a2}$ = -3e3    & $q_2$ = 0.55mL/s   \\
    p$K_1$ = 6.35   & $W_{b2}$ = 3e3     & $q_3$ = 15.6mL/s   \\
    p$K_2$ = 10.25  & $W_{a3}$ = 3.05e3  & $q_4$ = 32.8mL/s   \\
    $h_1$ = 14 cm   & $W_{b3}$ = 5e1     & $A_1$ = 207 $cm^2$ \\
    pH = 7.0        & $W_{a4}$ = -4.32e2 & \\
    \bottomrule
    \label{tab:pH_parameters}
    \end{tabular}
\end{table}
\end{extendedonly}
The input of the system is constrained to be in the range $u \in [11.2, 17.2]$.
In the following we assume that the only information available for identification and control purposes is the input-output data.
This data-set was generated by simulating \eqref{eq:example:ph_process} as discussed below.

\subsection{Identification}
To build the data-set, the simulated process is excited by a Multilevel Pseudo-Random Signal (MPRS) with a sampling time of $\tau_s = 10 s$, so that at least $30$ input-output data points are collected during the settling time of each input step.
White noise is added both to the input and output with standard deviations of $10^{-3}$ and $7.5\cdot 10^{-3}$, respectively. 
The resulting training set consists of an experiment of $5060$ input-output pairs $(u, y)$, corresponding to $14$ hours of operation. 
In order to facilitate training and satisfy the assumptions made in Section \ref{sec:model}, both the input and output need to be normalized, so that $y \in [-1, 1]$ and $u \in [-1, 1]$.

\begin{shortonly}
    \begin{figure}[t]
        \centering
        \includegraphics[width=0.5\linewidth]{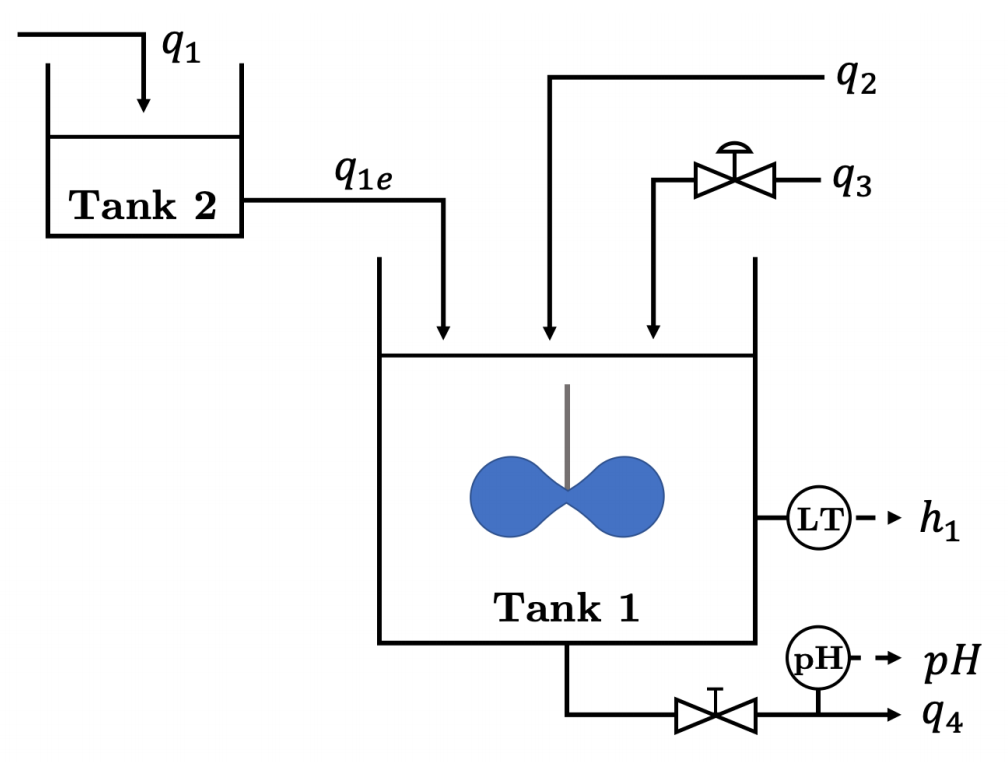}
        \caption{A representation of the pH neutralization process.}
        \label{fig:pH_scheme}
    \end{figure}
\end{shortonly}

The GRU network described in Section \ref{sec:model} is used to identify the unknown plant.
In light of the recurrent nature of the GRU, the so-called Truncated Back-Propagation Through Time (T-BPTT) is adopted, which consists in splitting the experiment in $N_s$ partially overlapping sequences of length $T_s=1000$, denoted by $(\bm{u}^{\{i\}}, \bm{y}^{\{i\}}) $. More specifically, the $i$-th input and output sequence are obtained as follows
\begin{equation*}
    \scalemath{0.85}{
    \begin{aligned}
        \bm{u}^{\{i\}} = \{ u(i \tau +1), ..., u(i \tau + L_s) \}, \, \bm{y}^{\{i\}} = \{ y(i \tau +1), ..., y(i \tau + L_s) \},
    \end{aligned}}
\end{equation*}
where $\tau = 5$ is the number of non-overlapping samples. 
\ifbool{extended}{In Figure \ref{fig:training_sequences} some of these input and output sequences are depicted.}{}

At each training step, a random subset $\mathcal{I}$ of\ifbool{extended}{}{ these} sequences is extracted. 
For each $i \in \mathcal{I}$, a random initial state $\bar{x}$ of the network is generated, and the network \eqref{eq:model:gru} is simulated in open loop with the input sequence $\bm{u}^{\{i\}}$.
This operation artificially augments the training set and causes the network to minimize the $T_s$-step-ahead prediction error, which is desirable to ensure that the prediction model of the NMPC is accurate throughout the horizon.

The loss function $L$ adopted for training is the Mean-Squared Error (MSE) between the measured output sequence $\bm{y}^{\{i\}}$ and the output trajectory predicted by the network, $y(k, \bar{x}, \bm{u}^{\{i\}})$, obtained initializing \eqref{eq:model:gru} in the random state $\bar{x}$ and feeding it with the input sequence $\bm{u}^{\{ i\}}$.
A washout period of $T_w$ steps is considered, meaning that the simulation error of the first $T_w$ steps is not penalized, to accommodate the network's initial transitory due to the random initialization.
Hence, the adopted loss functions is
\begin{equation} \label{eq:example:loss}
\scalemath{0.875}{
	L = \sum_{i\in \mathcal{I}} \bigg[ \frac{1}{T_s - T_{w}} \sum_{k=T_w}^{T_s} \big\| y(k, \bar{x}, \bm{u}^{\{i\}}) - \bm{y}^{\{i\}}(k) \big\|^2 \bigg] + \rho(\nu),  }
\end{equation}
where $\rho(\nu)$ is a term penalizing the violation of the $\delta$ISS condition \eqref{eq:model:deltaiss_condition}, which reads as
\begin{equation}\label{eq:residual_dISS}
\scalemath{0.875}{
	\nu = \| U_r \|_\infty \Big(\frac{1}{4} \| U_f \|_\infty + \bar{\sigma}_f \Big) + \frac{1}{4} \frac{1 + \bar{\phi}_r}{1 - \bar{\sigma}_z} \| U_z \|_\infty - 1.}
\end{equation}
Steering $\nu$ to negative values one obtains, according to Theorem \ref{thm:deltaiss}, a provably $\delta$ISS network.
A strictly increasing $\rho(\nu)$ function, such as a piece-wise linear function, should be hence adopted \citep{bonassi2020stability}.
Here we consider
\begin{equation}
\scalemath{0.875}{
    \rho(\nu) = \rho^+ \max(0, \nu) + \rho^- \min(0, \nu),}
\end{equation}
with $\rho^+ = 10^{-2}$ and $\rho^- = 10^{-6}$.

\begin{extendedonly}
    \begin{figure}[t]
    \centering
    \includegraphics[width=0.95\linewidth]{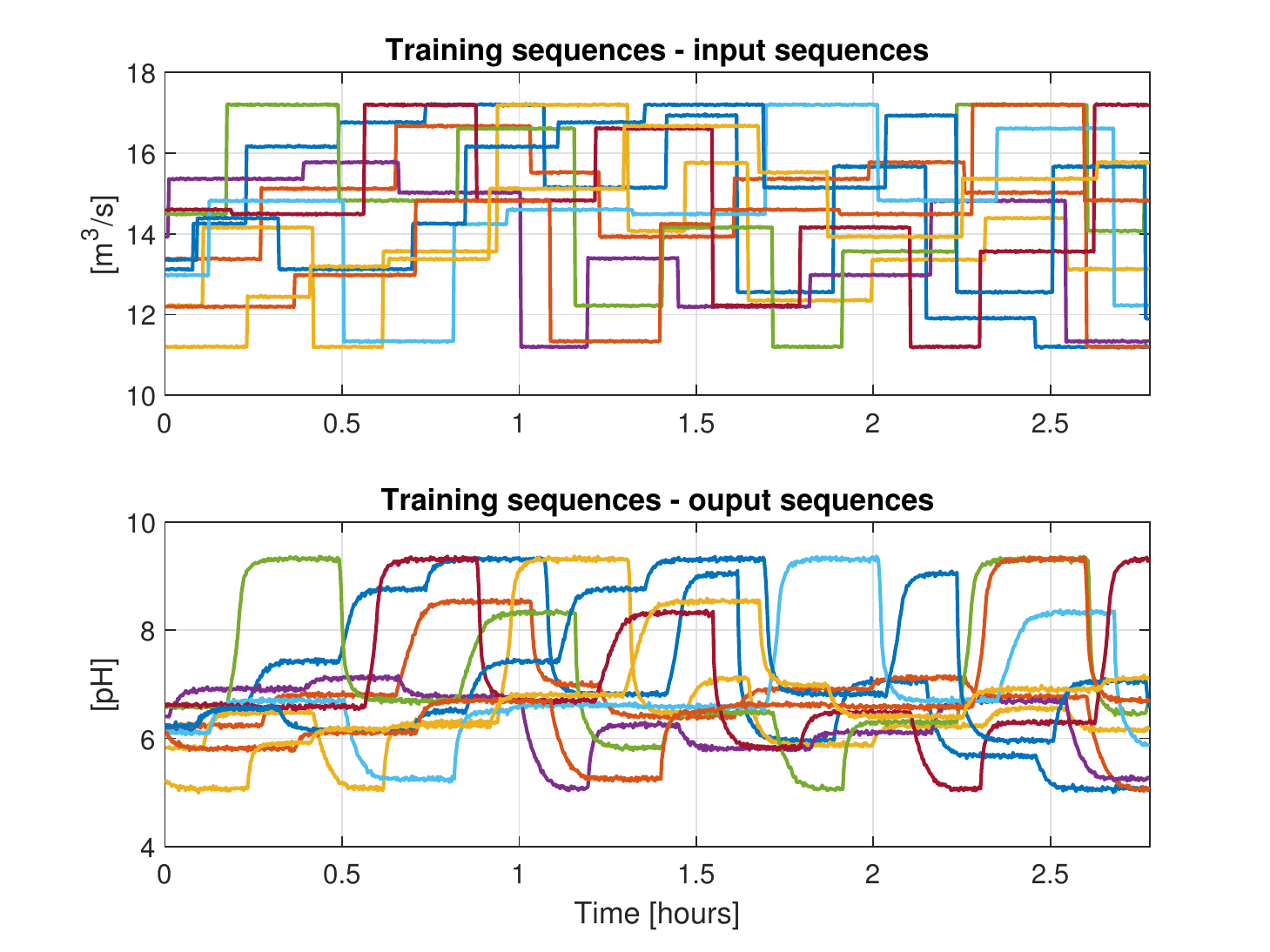}
    \caption{Samples of input and output sequences used for network's training.}
    \label{fig:training_sequences}
\end{figure}
\end{extendedonly}

\begin{figure}[t]
    \centering
    \includegraphics[width=0.95\linewidth]{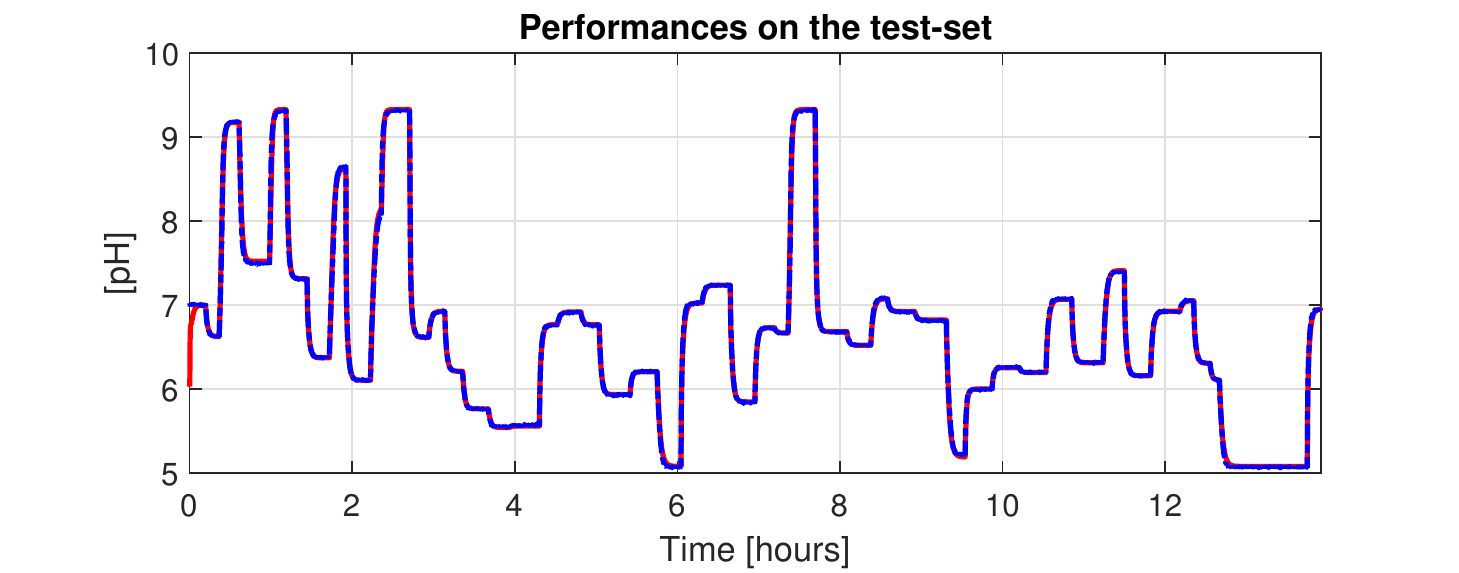}
    \caption{Performances of the GRU network on the test set:  network output (red continuous line) versus the real measured output (blue dashed-dotted line).}
    \label{fig:gru_testdata}
\end{figure}

The implemented GRU network \eqref{eq:model:gru} has $n = 10$ neurons, and it has been trained for $200$ epochs, using the ADAM optimization algorithm to minimize the loss function \eqref{eq:example:loss}.

The training has been carried out in Python 3.7 with the PyTorch 1.5 deep learning library. 
At the end of the training procedure, the performances of the network have been tested on an independent test set, obtaining the results depicted in Figure \ref{fig:gru_testdata}.
The MSE on the test set is approximately $10^{-3}$.
The accuracy is also quantified by the FIT index $[\%]$, defined as
\begin{equation*}
\scalemath{0.85}{
    \text{FIT} = 100\left(1 - \frac{\| \bm{y}^{\{ts\}} - \bm{y} \|}{\| \bm{y}^{\{ts\}} - y_{avg} \|} \right)}
\end{equation*}
where $\bm{y}^{\{ ts \}}$ is the test set output, $y_{avg}$ is its average, and $\bm{y}$ is the output of the network, initialized in the random state $\bar{x}$ and fed by the test input sequence $\bm{u}^{\{ts\}}$. 
The trained network scores $\text{FIT} = 97.34\%$, which confirms yet again the remarkable modeling capability of GRU networks. 
Moreover, at the end of training, the residual \eqref{eq:residual_dISS} is negative, $\nu=-0.0284$, meaning that the condition for the $\delta$ISS is successfully enforced.

\subsection{Control}
This section aims to test the ability of the NMPC controller to robustly track constant set-points. 
The state estimates are provided by a detector $\mathcal{O}_a$, defined as \eqref{eq:control:detector}, whose gains are determined by the solution of the optimization problem \eqref{eq:control:observer_synthesis}. 
The control and prediction horizons are selected as $N_c=20$ and $N_p=75$, while the adopted weighting matrices are $Q=I$ and $R=1$. 
For the computation of the terminal region, we take $\tilde{Q}=10 \, I$ and $\gamma = 0.01$. 
The terminal cost $V_f$ \eqref{eq:control:terminal_cost} is adopted, with $N_f = 10^3$.
A moving average filter is applied to the reference signal to smooth out the transitions between different references.

In order to test the controller robustness, three different disturbances are also applied on the system. 
The first disturbance, applied between $t=0.5 h$ and $ t=1.5h$, is a constant additive disturbance on the plant output with amplitude $w_{out} = -0.5$, which represents about $10\%$ of the output range.
Then, from $t=2.5 h$ to $t=3.5 h$, the flow $q_2$ is changed from $0.55$ to $0.4$. 
Eventually, a constant additive input disturbance is introduced from  $t=4.5 h$ to $t=5.5 h$, with amplitude $w_{in}=0.6$, corresponding to $10\%$ of the input range. 

\begin{figure}[t]
    \centering
    \includegraphics[width=0.9\linewidth]{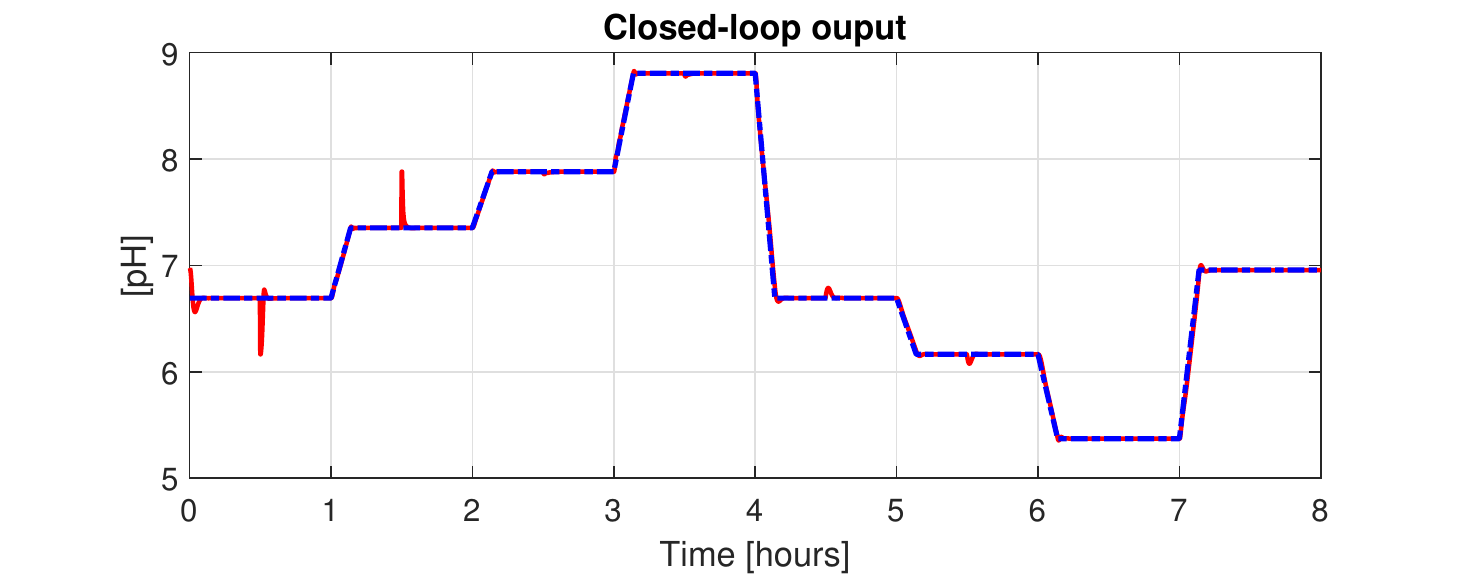}
    \vspace{-1mm}
    \caption{Closed-loop performances: plant output (red solid line) versus reference (blue dashed-dotted line).}
    \label{fig:closed_loop}
\end{figure}

\begin{figure}[t]
    \centering
    \includegraphics[width=0.95\linewidth]{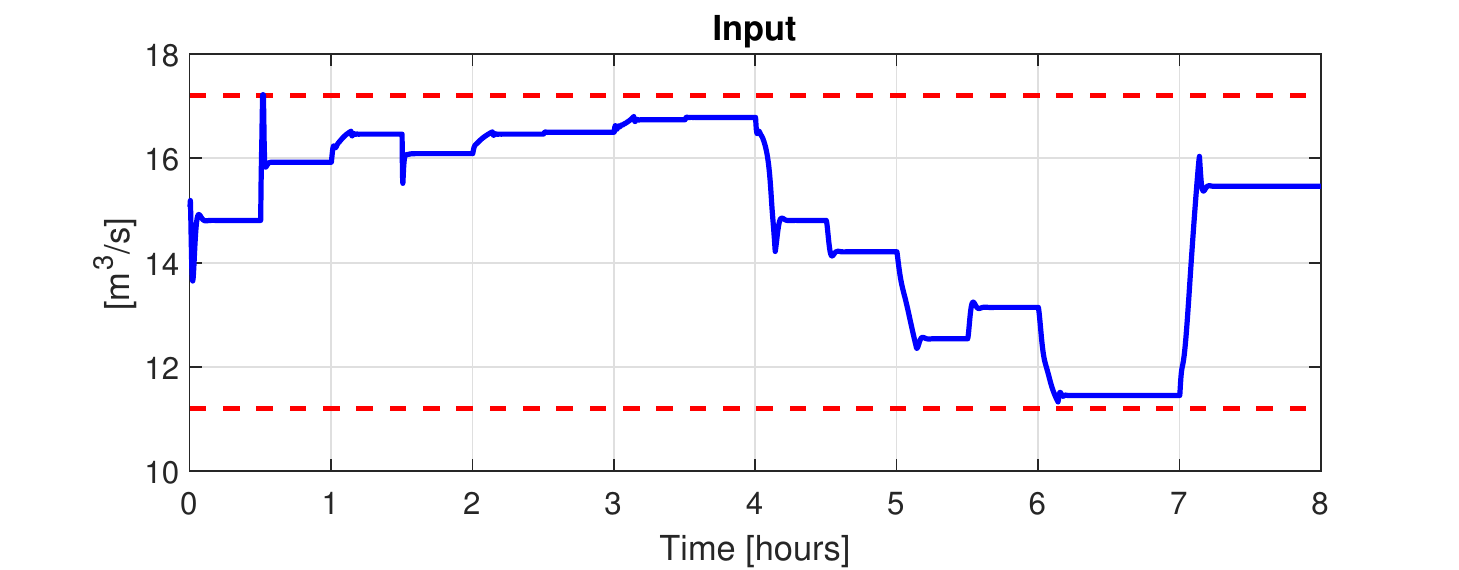}
    \vspace{-1mm}
    \caption{The denormalized input $u = v + \xi$ (blue line) is bounded within its minimum and maximum values.}
    \label{fig:input}
\end{figure}

The output of the closed-loop system is shown Figure \ref{fig:closed_loop}\ifbool{extended}{, and the corresponding tracking error is depicted in Figure \ref{fig:tracking_error}.}{.}
It is apparent that the controller successfully compensates the modeling errors and external unknown disturbances, achieving offset-free tracking, while fulfilling the constraints on the input variable $u$, see Figure \ref{fig:input}. 
Even though the control system is based on the nominal system, the performance translates well into the real plant, owing to integral action and accurate modeling.

\begin{extendedonly}
    \begin{figure}[t]
    \centering
    \includegraphics[width=0.9\linewidth]{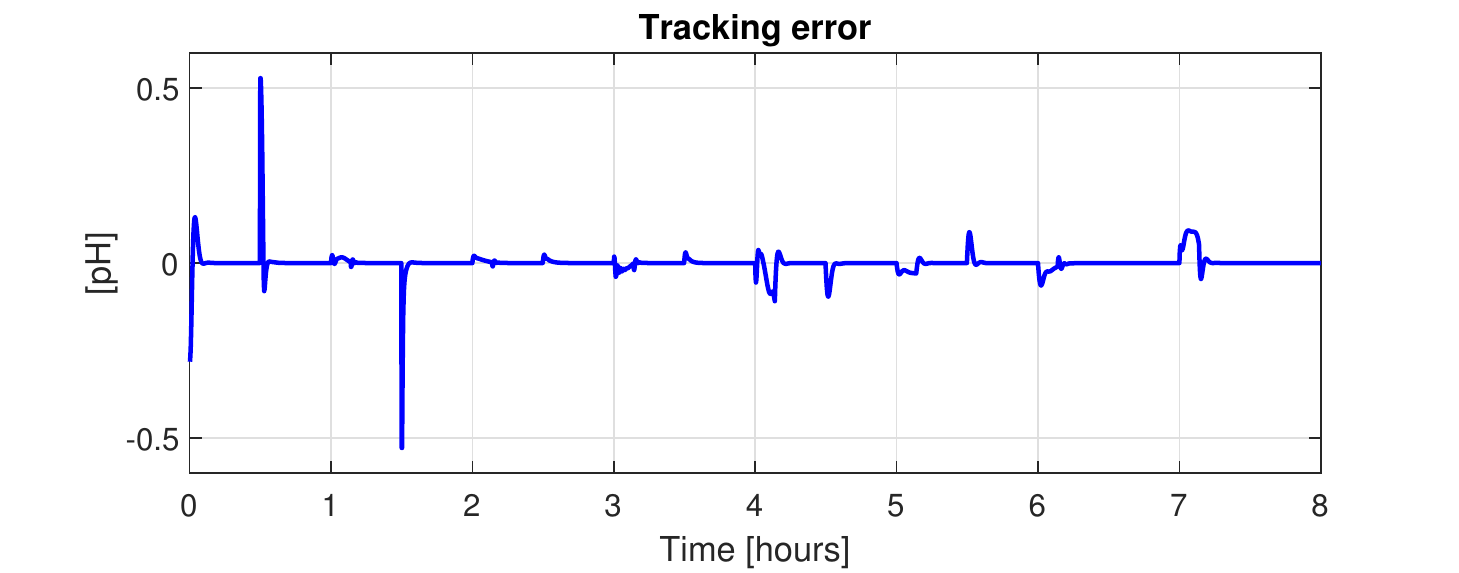}
    \vspace{-1mm}
    \caption{Closed-loop tracking error.}
    \label{fig:tracking_error}
    \end{figure}
\end{extendedonly}

\begin{rem}
    If the set-point changes, the equilibrium associated with the new set-point must be re-computed, as well as (in principle) the auxiliary control law and the terminal set $\Omega_\omega$. 
    While this computation may be time consuming for on-line operation, $\Omega_{\omega}$ can be computed in advance for any target equilibrium $(x_a^0, 0, y_a^0)$ by applying \eqref{eq:control:terminal}.
\end{rem}

\section{Conclusions}
In this paper we addressed the task of stable offset-free tracking of constant references for nonlinear stable dynamical systems learned by Gated Recurrent Units (GRU).
To this end, a GRU network was firstly trained to identify the unknown plant, ensuring the stability of the model by fulfilling the stability conditions provided in \cite{bonassi2020stability}.
Then, to ensure offset-free tracking, the model was augmented with an integrator, and a converging state observer was setup for the augmented model.
The control strategy provided by \cite{magni2001output}, guaranteeing stable offset-free tracking of constant references, was applied. 
The proposed approach was tested on the \emph{pH} neutralization process benchmark, witnessing remarkable performances and good robustness to disturbances.

\bibliography{GRU_Offset_Free}

\begin{thebibliography}{18}
\providecommand{\natexlab}[1]{#1}
\providecommand{\url}[1]{\texttt{#1}}
\providecommand{\urlprefix}{URL }
\expandafter\ifx\csname urlstyle\endcsname\relax
  \providecommand{\doi}[1]{doi:\discretionary{}{}{}#1}\else
  \providecommand{\doi}{doi:\discretionary{}{}{}\begingroup
  \urlstyle{rm}\Url}\fi

\bibitem[{Bonassi et~al.(2020{\natexlab{a}})Bonassi, Farina, and
  Scattolini}]{bonassi2020stability}
Bonassi, F., Farina, M., and Scattolini, R. (2020{\natexlab{a}}).
\newblock On the stability properties of gated recurrent units neural networks.
\newblock \emph{arXiv preprint arXiv:2011.06806}.

\bibitem[{Bonassi et~al.(2020{\natexlab{b}})Bonassi, Terzi, Farina, and
  Scattolini}]{bonassi2019lstm}
Bonassi, F., Terzi, E., Farina, M., and Scattolini, R. (2020{\natexlab{b}}).
\newblock {LSTM} neural networks: Input to state stability and probabilistic
  safety verification.
\newblock In \emph{Learning for Dynamics and Control}, 85--94.

\bibitem[{Chung et~al.(2014)Chung, Gulcehre, Cho, and
  Bengio}]{chung2014empirical}
Chung, J., Gulcehre, C., Cho, K., and Bengio, Y. (2014).
\newblock Empirical evaluation of gated recurrent neural networks on sequence
  modeling.
\newblock \emph{arXiv preprint arXiv:1412.3555}.

\bibitem[{De~Nicolao et~al.(1997)De~Nicolao, Magni, and
  Scattolini}]{de1997narx}
De~Nicolao, G., Magni, L., and Scattolini, R. (1997).
\newblock Stabilizing predictive control of nonlinear arx models.
\newblock \emph{Automatica}, 33(9), 1691--1697.

\bibitem[{Forgione and Piga(2020)}]{forgione2020model}
Forgione, M. and Piga, D. (2020).
\newblock Model structures and fitting criteria for system identification with
  neural networks.
\newblock In \emph{2020 IEEE 14th International Conference on Application of
  Information and Communication Technologies (AICT)}, 1--6. IEEE.

\bibitem[{{Hall} and {Seborg}(1989)}]{pHbenchmark}
{Hall}, R.C. and {Seborg}, D.E. (1989).
\newblock Modelling and self-tuning control of a multivariable ph
  neutralization process part i: Modelling and multiloop control.
\newblock In \emph{1989 American Control Conference}, 1822--1827.

\bibitem[{Hochreiter and Schmidhuber(1997)}]{hochreiter1997long}
Hochreiter, S. and Schmidhuber, J. (1997).
\newblock Long short-term memory.
\newblock \emph{Neural computation}, 9(8), 1735--1780.

\bibitem[{Lanzetti et~al.(2019)}]{lanzetti2019recurrent}
Lanzetti, N. et~al. (2019).
\newblock Recurrent neural network based {MPC} for process industries.
\newblock In \emph{2019 18th European Control Conference (ECC)}, 1005--1010.
  IEEE.

\bibitem[{Magni et~al.(2001{\natexlab{a}})Magni, De~Nicolao, Magnani, and
  Scattolini}]{magni2001stabilizing}
Magni, L., De~Nicolao, G., Magnani, L., and Scattolini, R.
  (2001{\natexlab{a}}).
\newblock A stabilizing model-based predictive control algorithm for nonlinear
  systems.
\newblock \emph{Automatica}, 37(9), 1351--1362.

\bibitem[{Magni et~al.(2001{\natexlab{b}})Magni, De~Nicolao, and
  Scattolini}]{magni2001bookchapter}
Magni, L., De~Nicolao, G., and Scattolini, R. (2001{\natexlab{b}}).
\newblock Model predictive control: output feedback and tracking of nonlinear
  systems.
\newblock In \emph{Non-linear Predictive Control: theory and practice},
  chapter~3, 61--80. Institution of Engineering and Technology.

\bibitem[{Magni et~al.(2001{\natexlab{c}})Magni, De~Nicolao, and
  Scattolini}]{magni2001output}
Magni, L., De~Nicolao, G., and Scattolini, R. (2001{\natexlab{c}}).
\newblock Output feedback and tracking of nonlinear systems with model
  predictive control.
\newblock \emph{Automatica}, 37(10), 1601--1607.

\bibitem[{Morari and Maeder(2012)}]{morari2012}
Morari, M. and Maeder, U. (2012).
\newblock Nonlinear offset-free model predictive control.
\newblock \emph{Automatica}, 48(9), 2059--2067.

\bibitem[{Ohno(1978)}]{ohno1978new}
Ohno, K. (1978).
\newblock A new approach to differential dynamic programming for discrete time
  systems.
\newblock \emph{IEEE Transactions on Automatic Control}, 23(1), 37--47.

\bibitem[{Pannocchia et~al.(2015)Pannocchia, Gabiccini, and
  Artoni}]{pannocchia2015offset}
Pannocchia, G., Gabiccini, M., and Artoni, A. (2015).
\newblock Offset-free mpc explained: novelties, subtleties, and applications.
\newblock \emph{IFAC-PapersOnLine}, 48(23), 342--351.

\bibitem[{Stipanovi{\'c} et~al.(2020)Stipanovi{\'c}, Kapetina, Rapai{\'c}, and
  Murmann}]{stipanovic2020stability}
Stipanovi{\'c}, D.M., Kapetina, M.N., Rapai{\'c}, M.R., and Murmann, B. (2020).
\newblock Stability of gated recurrent unit neural networks: Convex combination
  formulation approach.
\newblock \emph{Journal of Optimization Theory and Applications}, 1--16.

\bibitem[{Terzi et~al.(2020)Terzi, Bonetti, Saccani, Farina, Fagiano, and
  Scattolini}]{terzi2020learning}
Terzi, E., Bonetti, T., Saccani, D., Farina, M., Fagiano, L., and Scattolini,
  R. (2020).
\newblock Learning-based predictive control of the cooling system of a large
  business centre.
\newblock \emph{Control Engineering Practice}, 97, 104348.

\bibitem[{Terzi et~al.(2021)Terzi, Bonassi, Farina, and
  Scattolini}]{terzi2021learning}
Terzi, E., Bonassi, F., Farina, M., and Scattolini, R. (2021).
\newblock Learning model predictive control with long short-term memory
  networks.
\newblock \emph{International Journal of Robust and Nonlinear Control}, 1--20.
\newblock \doi{10.1002/rnc.5519}.

\bibitem[{Wong et~al.(2018)}]{wong2018recurrent}
Wong, W. et~al. (2018).
\newblock Recurrent neural network-based model predictive control for
  continuous pharmaceutical manufacturing.
\newblock \emph{Mathematics}, 6(11), 242.

\end{thebibliography}

\begin{extendedonly}
\appendix
\section{Proofs}

\textbf{Proof of Theorem \ref{thm:observer}}
This proof is organized as follows: first, we show that the Schur stability of $A_\delta$ implies that the state estimate $\hat{x}_a$ nominally converges to the true state $x_a$. 
Then, we show that this implies that $\mathcal{O}_a$ is a weak detector of the augmented system $\Sigma_a$.

Let $e_x = x - \hat{x}$ and $e_\xi = \xi - \hat{\xi}$. Then, the $j$-th component of $e_x$ reads as
\begin{equation}
    \begin{aligned}
    e_{x j}^+ =& z_j x_j + (1-z_j) \phi(W_r(v + \xi) + U_r f \circ x + b_r)_j \\
    & - \hat{z}_j \hat{x}_j - (1-\hat{z}_j) \phi(W_r(v + \hat{\xi}) + U_r \hat{f} \circ \hat{x} + b_r)_j.
    \end{aligned}
\end{equation}
Taking the absolute value, being $z_j \in (0, 1)$, we obtain
\begin{equation} \label{eq:proofs:thm2:absv}
    \begin{aligned}
    \lvert e_{xj}^+ \lvert \leq& z_j \lvert x_j - \hat{x}_j\lvert + \lvert z_j - \hat{z}_j\lvert \, \lvert \hat{x}_j \lvert \\
    &+  \lvert z_j - \hat{z}_j \lvert \, \big\lvert \phi(W_r(v + \hat{\xi}) + U_r \hat{f} \circ \hat{x} + b_r)_j \big\lvert \\
    & + (1-z_j) \, \big\lvert \phi(W_r(v + \xi) + U_r f \circ x + b_r)_j \\
    & \quad\qquad\qquad - \phi(W_r(v + \hat{\xi}) + U_r \hat{f} \circ \hat{x} + b_r)_j\big\lvert.
    \end{aligned}
\end{equation}
\begin{subequations} \label{eq:proofs:thm2:steps}
Since $\sigma$ is Lipschitz-continuous with Lipschitz constant $\frac{1}{4}$, and since $\lvert z_j - \hat{z}_j \lvert \leq \| z_j - \hat{z}_j \|_\infty $, it holds that
\begin{equation}
\scalemath{0.95}{
    \begin{aligned}
    \lvert z_j - \hat{z}_j \lvert \leq \frac{1}{4} \big( \| W_z - L_{z \xi}  \|_\infty \| e_\xi \|_\infty + \| U_z - L_{z y} U_o \|_\infty \| e_x \|_\infty \big).
    \end{aligned}}
\end{equation}
Owing to Lemma \ref{lemma:invset}, both $\| x\|_\infty \leq 1$ and $\| \hat{x}\|_\infty \leq 1$, thus
\begin{equation}
    \big\lvert \phi(W_r (v + \hat{\xi}) + U_r \hat{f} \circ \hat{x} + b_r )_j \big\lvert \leq \bar{\phi}_r,
\end{equation}
where $\bar{\phi}_r$ is defined as in \eqref{eq:model:sigma_bar}.
Concerning the last term of \eqref{eq:proofs:thm2:absv}, in light of the Lipschitzianity of $\phi$,
\begin{equation}
\scalemath{0.825}{
    \begin{aligned}
    &\big\lvert \phi(W_r(v + \xi) + U_r f \circ x + b_r)_j - \phi(W_r(v + \hat{\xi}) + U_r \hat{f} \circ \hat{x} + b_r)_j\big\lvert \\
    & \leq \| W_r \|_\infty \| e_\xi \|_\infty + \|U_r \|_\infty \big\| ( f - \hat{f}) \circ \hat{x} + f \circ (x - \hat{x}) \big\|_\infty \\
    & \leq \| W_r \|_\infty \| e_\xi \|_\infty + \|U_r \|_\infty  \Big( \frac{1}{4} \| W_f - L_{f \xi} \|_\infty  \|e_\xi \|_\infty \\
    &\qquad\qquad + \frac{1}{4} \| U_f - L_{f x} U_o \|_\infty \| e_x \|_\infty + \bar{\sigma}_f \| e_x \|_\infty \Big),
    \end{aligned}}
\end{equation}
where $\bar{\sigma}_f$ is defined as in \eqref{eq:model:sigma_bar}.
\end{subequations}
Combining \eqref{eq:proofs:thm2:absv} and \eqref{eq:proofs:thm2:steps} one gets
\begin{equation}
\scalemath{0.85}{
    \begin{aligned}
    \lvert e_{x j}^+ \lvert \leq & \bigg[ z_j + (1-z_j) \| U_r \|_\infty \Big( \frac{1}{4} \| U_f - L_{fy} U_o \|_\infty + \bar{\sigma}_f \Big) \\
    & + \frac{1}{4} \| U_z \|_\infty (1 + \bar{\phi}_r) \bigg] \| e_{x j} \|_\infty + \alpha \| e_{\xi} \|_\infty,
    \end{aligned}}
\end{equation}
where $\alpha$ is defined by \eqref{eq:control:observer_theorem:a12}. 
Then, if \eqref{eq:control:observer_theorem:xx} holds, the previous inequality becomes
\begin{equation} \label{eq:proofs:thm2:e_x}
    \| e_x^+ \|_\infty \leq (1-\delta) \| e_x \|_\infty + \alpha \|e_\xi \|_\infty.
\end{equation}

Concerning the observation error $e_\xi = \xi - \hat{\xi}$,
\begin{equation}\label{eq:proofs:thm2:e_xi}
    e_\xi^+ = \| I + L_{\xi y} U_o \|_\infty \| e_x \|_\infty + \| I - L_{\xi \xi} \|_\infty \| e_\xi \|_\infty
\end{equation}
Combining \eqref{eq:proofs:thm2:e_x} and \eqref{eq:proofs:thm2:e_xi} it follows that
\begin{equation} \label{eq:proofs:thm2:error}
    \begin{bmatrix}
    \| e_x^+ \|_\infty \\ \| e_\xi^+ \|_\infty 
    \end{bmatrix} = \underbrace{\begin{bmatrix}
    1- \delta & & \alpha \\
    \| U_o \|_\infty \| I + L_{\xi y}\|_\infty &\,\,\,& \| I - L_{\xi \xi} \|_\infty 
    \end{bmatrix}}_{A_{\delta}} \begin{bmatrix}
    \| e_x \|_\infty \\ \| e_\xi \|_\infty 
    \end{bmatrix},
\end{equation}
and therefore the Schur stability of $A_\delta$ implies that the state observation error converges to zero.

Being $A_\delta$ Schur stable, there exist a symmetric positive definite matrix $P$ solving the Lyapunov equation $A_\delta^\prime P A_\delta - P \preccurlyeq - I$.
Let us consider
\begin{equation}
\scalemath{0.9}{
    V_e(x_a, \hat{x}_a) = \begin{bmatrix}
    \| e_x \|_\infty \\ \| e_\xi \|_\infty 
    \end{bmatrix}^\prime P \begin{bmatrix}
    \| e_x \|_\infty \\ \| e_\xi \|_\infty 
    \end{bmatrix} = \left\| \begin{bmatrix}
    \| e_x \|_\infty \\ \| e_\xi \|_\infty 
    \end{bmatrix} \right\| ^2_P}
\end{equation}
as Lyapunov function candidate.
We recall that 
\begin{equation*}
\scalemath{0.9}{
    \lambda_m(P) \| v \|^2 \leq \| v \|_P^2 \leq \lambda_M(P) \| v \|_P^2,}
\end{equation*}
where $\lambda_m(P)$ and $\lambda_M(P)$ are the minimum and maximum singular values of $P$, respectively, and that $\frac{1}{\sqrt{n}} \| v \| \leq \| v \|_\infty \leq \| v \|$. 
Since $\| [ \| e_x \|_\infty, \| e_\xi \|_\infty ]^\prime \|_\infty = \| x_a - \hat{x}_a \|_\infty$ one can obtain
\begin{equation}
\scalemath{0.9}{
\left\| \begin{bmatrix}
    \| e_x \|_\infty \\ \| e_\xi \|_\infty 
    \end{bmatrix} \right\| ^2_P \leq  2 \lambda_M(P) \| x_a - \hat{x}_a \|^2_\infty \leq 2 \lambda_M(P) \| x_a - \hat{x}_a \|^2}
\end{equation}
and
\begin{equation}
\scalemath{0.9}{
    \left\| \begin{bmatrix}
    \| e_x \|_\infty \\ \| e_\xi \|_\infty 
    \end{bmatrix} \right\| ^2_P \geq \lambda_m(P) \| x_a - \hat{x}_a \|_\infty^2 \geq  \frac{1}{n+p} \lambda_m(P) \| x_a - \hat{x}_a\|^2.}
\end{equation}
The Lyapunov function candidate is thus bounded by
\begin{equation}
\scalemath{0.9}{
    \frac{1}{n+p} \lambda_m(P) \left\| x_a - \hat{x}_a \right\|^2 \leq V_e(x_a, \hat{x}_a) \leq 2 \lambda_M(P) \left\| x_a - \hat{x}_a \right\|^2}.
\end{equation}
Using the same arguments it can be shown that
\begin{equation}
    V(x_a^+, \hat{x}_a^+) - V(x_a, \hat{x}_a) \leq - \frac{1}{n+p} \left\| x_a - \hat{x}_a \right\|^2.
\end{equation}
Thus, by Definition \ref{def:detector}, the augmented system \eqref{eq:control:augmented_simple} is weakly detectable and  $\mathcal{O}_a$ is its weak detector. $\hfill \blacksquare$

\smallskip
\textbf{Proof of Proposition \ref{prop:obsv_tuning}}

In light of the Jury criterion, matrix $A_\delta$ defined as  \eqref{eq:control:observer_theorem:A} is Schur stable if and only if
\begin{equation} \label{eq:proofs:prop1:jury}
    -1+A_{\delta11}+A_{\delta22} < A_{\delta11} A_{\delta22} - A_{\delta12} A_{\delta21} < 1, 
\end{equation}
where $A_{\delta ij}$ denotes the element of $A_\delta$ in position $(i,j)$.
Then, \eqref{eq:proofs:prop1:jury} implies
\begin{equation}
\scalemath{0.9}{
    \begin{aligned}
    & \delta + \| I - L_{\xi \xi} \|_\infty < (1-\delta) \| I - L_{\xi \xi} \|_\infty  - \alpha \| U_o \|_\infty \| I + L_{\xi y} \|_\infty, \\
    &(1-\delta) \| I - L_{\xi \xi} \|_\infty  - \alpha \| U_o \|_\infty \| I + L_{\xi y} \|_\infty < 1,
    \end{aligned}}
\end{equation}
from which the constraints of the optimization problem \eqref{eq:control:observer_synthesis} can be easily derived. 
Therefore, provided that these constraints are satisfied, the Jury criterion ensures the Schur stability of $A_\delta$, and $\mathcal{O}_a$ fulfills the assumptions of Theorem  \ref{thm:observer}, meaning that it is guaranteed to be a weak detector of the augmented system.
$\hfill \blacksquare$

\smallskip
\textbf{Proof of Corollary \ref{cor:trivial_observer}}

If the model \eqref{eq:model:gru} satisfies the $\delta$ISS sufficient condition \eqref{eq:model:deltaiss_condition}, taking $L_{fy} = 0_{n, p}$ and $L_{zy} = 0_{n, p}$, the variable $\delta$ reads as
\begin{equation}
\scalemath{0.9}{
    \delta = 1 - \| U_r \|_\infty \Big(\frac{1}{4} \| U_f \|_\infty + \bar{\sigma}_f \Big) - \frac{1}{4} \frac{1 + \bar{\phi}_r}{1 - \bar{\sigma}_z} \| U_z |_\infty > 0.}
\end{equation}
Taking $L_{z\xi} = W_z$ one has $\alpha = 0$, therefore the second constraint of \eqref{eq:control:observer_synthesis} becomes
\begin{equation}
    \delta ( 1- \| I - L_{\xi \xi} \|_\infty ) > 0,
\end{equation}
which is trivially satisfied if $L_{\xi \xi} = \lambda I_{p, p}$, with $\lambda \in (0, 1)$, since $\| I - L_{\xi \xi} \|_\infty < 1$.
Moreover, the third constraint becomes
\begin{equation}
    (1-\delta) \| I - L_{\xi \xi} \|_\infty < 1,
\end{equation}
which is satisfied since $\delta \in (0, 1)$. $\hfill \blacksquare$
\end{extendedonly}

\end{document}